\RequirePackage[l2tabu, orthodox]{nag}
\RequirePackage{fixltx2e}
\documentclass[twocolumn, 10pt, aps, superscriptaddress, floatfix, showpacs, pra, citeautoscript]{revtex4-1}
\usepackage{graphicx}
\usepackage{amsmath}
\usepackage{amssymb}
\usepackage{bm}
\usepackage[utf8]{inputenc}
\usepackage{xcolor}
\usepackage[colorlinks, citecolor={blue!50!black}, urlcolor={blue!50!black}, linkcolor={red!50!black}]{hyperref}
\usepackage{bookmark}
\usepackage{tabularx}
\usepackage{mathtools}
\usepackage{microtype}

\newcommand{\vt}[1]{\mathbf{#1}}

\newcommand{\co}[2]{#2}
\renewcommand{\paragraph}{\co}

\DeclarePairedDelimiter\abs{\lvert}{\rvert}%
\DeclarePairedDelimiter\norm{\lVert}{\rVert}%
\makeatletter
\let\oldabs\abs
\def\abs{\@ifstar{\oldabs}{\oldabs*}}
\let\oldnorm\norm
\def\norm{\@ifstar{\oldnorm}{\oldnorm*}}
\makeatother

\newcommand{\bra}[1]{\langle #1|}
\newcommand{\ket}[1]{|#1\rangle}

\newcolumntype{L}[1]{>{\raggedright\arraybackslash}p{#1}}
\newcolumntype{C}[1]{>{\centering\arraybackslash}p{#1}}
\newcolumntype{R}[1]{>{\raggedleft\arraybackslash}p{#1}}

\begin{document}

\title{Momentum-space Landau levels in driven-dissipative cavity arrays}

\author{Andrei C. Berceanu}
\affiliation{Departamento de F\'isica Te\'orica de la Materia
Condensada \& Condensed Matter Physics Center (IFIMAC), Universidad
Aut\'onoma de Madrid, Madrid 28049, Spain}
\author{Hannah M. Price}
\affiliation{INO-CNR BEC Center and Dipartimento di Fisica,
Universit\`{a} di Trento, I-38123 Povo, Italy}
\author{Tomoki Ozawa}
\affiliation{INO-CNR BEC Center and Dipartimento di Fisica,
Universit\`{a} di Trento, I-38123 Povo, Italy}
\author{Iacopo Carusotto}
\affiliation{INO-CNR BEC Center and Dipartimento di Fisica,
Universit\`{a} di Trento, I-38123 Povo, Italy}

\begin{abstract}
  We theoretically study the driven-dissipative Harper-Hofstadter model on a
  2D square lattice in the presence of a weak harmonic trap. Without pumping and loss, the eigenstates of this system can be understood, in certain limits, as momentum-space toroidal Landau levels, where the Berry curvature, a geometrical property of an energy band, acts like a momentum-space magnetic field. We show that key features of these eigenstates can be observed in the  steady-state of the driven-dissipative system under a monochromatic coherent drive, and present a realistic proposal for an optical experiment using state-of-the-art coupled cavity arrays. We discuss how such spectroscopic measurements may be used to probe effects associated both with the off-diagonal elements of the matrix-valued Berry connection and with the synthetic magnetic gauge.
\end{abstract}

\maketitle

\section{Introduction}

A planar system of electrons in a
strong magnetic field is the archetypal model for studying phenomena such as the
integer and fractional quantum Hall effects. With recent advances in creating
synthetic gauge fields, however, new horizons have opened up for simulating such topological phases of matter also with neutral particles, such as photons~\cite{hafezi2014synthetic} or ultracold atoms~\cite{dalibardrmp2011, goldman_repprog_2014, Goldman_arxiv_2015}. 
Rather than simply replicating previous measurements, experiments with synthetic gauge fields allow for unprecedented access to properties such as the eigenstates or eigenspectrum, while the tunability and controllability of these experiments offer the prospect of simulating novel physics. 

In the presence of a (synthetic) gauge field, the eigenstates making up an energy band can have nontrivial geometrical properties, as encoded in the Berry connection and Berry curvature defined below~\cite{berry, xiao2010berryreview}. Understanding the geometry of eigenstates in a band is of great importance, not least because the integral of the Berry curvature over the 2D Brillouin zone (BZ) gives the first Chern number: the topological invariant responsible for the integer quantum Hall effect~\cite{thouless}. Consequently, there has been much work in recent years to develop new techniques with which to probe the properties of energy bands in photonics and ultracold gases. For example, the Berry curvature can be measured in the semiclassical dynamics of a wavepacket in an optical lattice~\cite{dudarev,1chang, price, cominotti, dauphin, aidelsburger2015measuring, jotzu2014experimental} or in photon transport in a cavity array~\cite{ozawa2014qhe}. In all these cases, the physics can be most naturally understood by recognising that the Berry curvature acts like a magnetic field in momentum space~\cite{berry, bliokh2005spin, PhysRevD.12.3845, cooper2012designing}. 

The analogy between Berry curvature and magnetism is most powerful when a geometrical energy band is subjected to an additional weak harmonic potential~\cite{price2014magnetic}. Then, in the effective momentum-space Hamiltonian, a harmonic potential acts like the kinetic energy of a particle in real space. Just as the physical momentum $\vt{p}-\vt{A}({\bf r})$ is the sum of the canonical momentum $\vt{p}$ and the magnetic vector potential $\vt{A} ({\bf r})$ in the magnetic Hamiltonian, so the physical position $\vt{r}+\mathcal{A}_{n,n}({\bf p})$, is given by the canonical position $\vt{r}$ and the Berry connection $\mathcal{A}_{n,n}({\bf p})$ of band $n$ in the effective Hamiltonian~\cite{adams1959energy,nagaosa, murakami2003dissipationless, bliokh2005spin, fujita, bliokh2005topological,gosselin2006semiclassical}. The Berry curvature, $\Omega_n ({\bf p})= \nabla \times \mathcal{A}_{n,n}({\bf p})$, is then like a momentum-space magnetic field. For certain models, this analogy leads to a clear analytical understanding of single-particle dynamics~\cite{price2014magnetic, ozawa2014momhh, price2015sporbit, Claassen_prl_2015}. In particular, we will focus on the small-flux limit of the Harper-Hofstadter Hamiltonian~\cite{harper1955magnetic,hofstadter1976butterfly}, which is a model that has recently been realized in a multitude of experimental configurations,
ranging from ultracold
gases~\cite{aidelsburger2013hh,miyake2013hh,mancini2015edge,stuhl2015edge},
solid state superlattices~\cite{dean2013hofstadter,yu2014hierarchy}
and silicon photonics~\cite{hafezi2013imaging} to classical systems
such as coupled pendula~\cite{susstrunk2015pendula} and oscillating
circuits~\cite{jia2013circuits}. As first shown in Ref.~\onlinecite{price2014magnetic}, the eigenstates of this model in the presence of a harmonic trap are toroidal Landau levels in momentum space. Not only would an observation of these states constitute the first exploration of analogue magnetic states in momentum space, but also the first experimental study of magnetism on a torus.  
 
While previous theoretical works on momentum-space Landau levels have focused on conservative dynamics~\cite{price2014magnetic, Claassen_prl_2015}, photonics systems naturally include driving and dissipation~\cite{carusotto2013fluids}. In this paper, we present a realistic experimental proposal for the observation of these states in a driven-dissipative 2D lattice of cavities, such as the array of coupled silicon ring resonators of Ref.~\cite{hafezi2013imaging}, where link resonators were used to simulate a synthetic gauge field for photons. In our proposal, we combine this set-up with a harmonic potential, introduced, for example, by a spatial modulation of the resonator size. We demonstrate numerically that the main features of momentum-space Landau levels will be observable spectroscopically in this system for realistic parameters. 

In this paper, we also emphasize how the inherent driving and dissipation in photonics can be a key advantage in probing properties that are otherwise inaccessible. 
Firstly, the spectroscopic measurements discussed here are sensitive to the absolute energy of a state. From this, we show how to extract the energy shift due to the off-diagonal matrix elements of the Berry connection $\mathcal{A}_{n,n'}({\bf p})$ relating eigenstates in different bands $n$ and $n'$. Only very recently has the first measurement of such effects been reported in ultracold atomic gases~\cite{Grusdt2014nonabelian,tracy2015arxiv}, and the approach used in this experiment would be difficult to apply in a photonics set-up. Our scheme may therefore be useful for the characterisation of energy bands in topologically-nontrivial photonic systems.

Secondly, since the photon steady-state depends on the overlap between the (observable) spatial amplitude profile of the drive and of the eigenstates~\cite{carusotto2013fluids}, the observables will depend on the phase of the eigenfunctions and thus on the specific synthetic magnetic gauge that is implemented in a given experimental realization of the Harper-Hofstadter Hamiltonian using a synthetic gauge field. We note that a related gauge-sensitivity has also recently been of much interest in ultracold gases in suitably designed time-of-flight experiments~\cite{kennedy2015bec,spielman2011gauge, spielman_gauge, tomoki2015nv}. Experiments on synthetic magnetic fields therefore present the opportunity of straightforwardly probing gauge-dependent physics. 

This paper is organized as follows: in Section \ref{sec:model} we introduce the trapped Harper-Hofstadter model, before reviewing how the eigenstates can be understood as momentum-space Landau levels in Section \ref{sec:eigenstates}. We discuss the breakdown of approximations in Section \ref{sec:berry-shift}, focusing on the energy shift from the off-diagonal matrix elements of the Berry connection. Then in Section \ref{sec:driven-dissipation}, we add driving and dissipation to the model. In Section \ref{sec:selection}, we show numerical results highlighting gauge-dependent effects, before presenting a viable proposal for a
photonics-based experiment in Section \ref{sec:experiment}. Finally, we draw conclusions in Section \ref{sec:conclusion}.

\section{Introduction to the trapped Harper-Hofstadter Model}
\label{sec:model}

In this paper, we study the Harper-Hofstadter Hamiltonian $\mathcal{H}_0$ in the presence of an external harmonic trap. The full tight-binding Hamiltonian $\mathcal{H}$ of this system is
\begin{eqnarray}
{\resizebox{.9\hsize}{!}{$\displaystyle{\mathcal{H}=\mathcal{H}_0+\frac{1}{2}\kappa
\sum_{m,n}\left[(m-m_0)^{2}+(n-n_0)^{2}\right]\hat{a}_{m,n}^{\dagger}\hat{a}_{m,n},}$}}  \qquad \label{eq:model}\\
{\resizebox{.9\hsize}{!}{$\displaystyle{\mathcal{H}_0=-J\sum_{m,n}(e^{i \phi_{m,n}^x}\hat{a}_{m+1,n}^{\dagger}\hat{a}_{m,n} 
+e^{i \phi_{m,n}^y}\hat{a}_{m,n+1}^{\dagger}\hat{a}_{m,n}) + \text{h.c.} }$}} \qquad\label{eq:hh_hamiltonian}
\end{eqnarray}
where $J$ is the real hopping amplitude and $\hat{a}_{m,n}^{\dagger}$ ($\hat{a}_{m,n}$) are the creation (annihilation) operators for a particle on a square lattice at site $(m,n)$. The harmonic trap is of strength $\kappa$ and is centered at a position $(m_0, n_0)$ which, in general, need not coincide with a lattice site. Throughout, the lattice spacing is set equal to one. 

In the Harper-Hofstadter model $\mathcal{H}_0$, the hopping phases $\phi = (\phi_{m,n}^x, \phi_{m,n}^y)$ are the Peierls phases gained by a charged particle hopping in the presence of a perpendicular magnetic field~\cite{harper1955magnetic,hofstadter1976butterfly}. The sum of the phases around a square plaquette of the lattice is therefore equal to $2\pi\alpha$, where $\alpha$ is the number of magnetic flux quanta through the plaquette (with $\hbar=e=1$). For neutral particles, such as photons, these phases can be imposed artificially to simulate the effects of magnetism, for example, by inserting link resonators into an array of silicon ring resonators as mentioned above~\cite{hafezi2013imaging}. 

Although the sum of phases around a plaquette is set by the external (synthetic) flux, the exact form of the hopping phases themselves depends on the choice of magnetic gauge. In the Landau gauge, for example, $\phi = (0, 2\pi\alpha m)$ such that only the hopping amplitude along one direction is modified. Conversely, in the symmetric gauge, $\phi = (-\pi\alpha n, \pi\alpha m)$ and so hopping terms along both $x$ and $y$ are affected, preserving the $C_4$ rotational invariance of the lattice. This gauge-dependence of the hopping phases is reflected in the spatial profile of the phase of an eigenstate of $\mathcal{H}$. In a photonics experiment, this phase is an observable quantity as the intensity response of a system to a given external driving is determined by the overlap of the spatial amplitude distribution of the pump with the eigenstates. Such experiments will therefore be sensitive to the synthetic magnetic gauge as we discuss in Section \ref{sec:experiment}.

\subsection{Toroidal Landau Levels in Momentum Space}\label{sec:eigenstates}

Having introduced the full Hamiltonian  in Eq. \refeq{eq:model}, we now review how the eigenstates of this model in an appropriate limit can be understood as toroidal Landau levels in momentum space~\cite{price2014magnetic}. Throughout the following discussion, we assume that the trap is centered at the origin $(m_0, n_0)= (0,0)$.  

We begin from the eigenstates of the Harper-Hofstadter model $\mathcal{H}_0\ket{\chi_{n,\vt{p}}} = E_n ({\bf p}) \ket{\chi_{n,\vt{p}}}$, where $E_n ({\bf p}) $ is the energy dispersion of band $n$ at crystal momentum $\vt{p}$. As the spatially-dependent hopping phases in $\mathcal{H}_0$ break translational invariance, new magnetic translation operators must be introduced to define a larger magnetic unit cell, containing an integer number of magnetic flux quanta~\cite{zak1964group, zak1964representations, 1chang}. Then translational symmetry is restored and Bloch's theorem can be applied to write the eigenstates as $\ket{\chi_{n,\vt{p}}} = \frac{1}{\sqrt{N}} e^{i\vt{p}\cdot \vt{r}} \ket{u_{n,\vt{p}}}$, where $\ket{u_{n,\vt{p}}}$ is the periodic Bloch function and $N$ is the number of lattice sites. Thanks to the new larger unit cell, the crystal momentum and the periodic Bloch functions here are defined in the smaller magnetic Brillouin zone (MBZ). For example, hereafter, we take the number of flux quanta per plaquette to be of the form $\alpha=1/q$, where $q$ is an integer. Then the magnetic unit cell can be chosen to be $q$ times larger than the original unit cell, while the MBZ is $q$ times smaller than the original BZ.

Adding the harmonic trap breaks all translational symmetry of the lattice, but we can use the eigenstates of $\mathcal{H}_0$ as a basis in which to expand the new wave function 
$\ket{\psi} = \sum_n\sum_{\vt{p}} \psi_n(\vt{p})
\ket{\chi_{n,\vt{p}}}$. Substituting this expansion into the full Schr\"{o}dinger equation
$i \partial_t \ket{\psi} = \mathcal{H} \ket{\psi}$, it can be shown that the expansion coefficients $\psi_n(\vt{p})$ satisfy~\cite{price2014magnetic}:
\begin{multline} 
  i \partial_t \psi_n(\vt{p}) = E_n(\vt{p}) + \frac{\kappa}{2}\sum_{n^{'},n^{''}}\left(\delta_{n,n^{'}}i \nabla_{\vt{p}} + \mathcal{A}_{n,n^{'}}(\vt{p})\right)\times \\ \times \left(\delta_{n^{'},n^{''}}i\nabla_{\vt{p}} + \mathcal{A}_{n^{'},n^{''}}(\vt{p})\right) \psi_{n^{''}}(\vt{p}) ,  \label{eq:first}
\end{multline}
where $\mathcal{A}_{n,n^{'}}(\vt{p}) = i\bra{u_{n,\vt{p}}}\nabla_{\vt{p}}\ket{u_{n^{'},\vt{p}}}$ is the matrix-valued Berry connection. 

To proceed, we consider the harmonic trap to be sufficiently weak compared to the bandgap that we can make a single-band approximation~\cite{price2014magnetic}. This assumes that only one coefficient $\psi_n$ is non-negligible and that the external trap does not significantly mix different energy bands. Then Eq. \ref{eq:first} reduces to
\begin{equation}
  i \partial_t \psi_n(\vt{p}) = \widetilde{\mathcal{H}} \psi_n(\vt{p}) ,
\end{equation}
where we have introduced the effective momentum-space Hamiltonian
\begin{equation}\label{eq:dual}
  \widetilde{\mathcal{H}} = \frac{\kappa}{2} [i\nabla_{\mathbf{p}} + \mathcal{A}_{n, n}(\mathbf{p})]^2 + E_n(\mathbf{p}) + \frac{\kappa}{2}\sum_{n^{'}\neq n} \abs{\mathcal{A}_{n,n^{'}}(\vt{p})}^2.
\end{equation}
For the moment we focus on the first two terms; we discuss the role of the last term, which comes from the off-diagonal matrix elements of the Berry connection, in detail in the next subsection.
As can be seen, there is a close analogy between the first two terms in the momentum-space Hamiltonian and that of a charged
particle in an electromagnetic field {\em in real space}:
\begin{eqnarray} \label{eq:mag}
\mathcal{H}'= \frac{\left[-i\nabla_{\vt{r}} - \vt{A}(\vt{r})\right]^2}{2M} +  \Phi({\bf r}). 
\end{eqnarray}
In this analogy, the role of the particle mass $M$ is played by $\kappa^{-1}$, while the scalar potential $ \Phi({\bf r})$ is replaced by the energy band dispersion $E_n(\mathbf{p})$ and the magnetic vector potential $\vt{A}(\vt{r})$ by the intra-band Berry connection $\mathcal{A}_{n, n}(\mathbf{p})$. We note that both the magnetic vector potential and the Berry connection are gauge-dependent quantities. We hereafter refer to the gauge choice for the Berry connection as the Berry gauge, and the gauge choice for a real-space magnetic vector potential as the magnetic gauge. From the Berry connection, we can also define the geometrical Berry curvature $\Omega_{n}(\mathbf{p}) =\nabla \times \mathcal{A}_{n, n}(\mathbf{p})$, which acts like a momentum-space magnetic field $B(\vt{r})$. 

The topology of momentum space also plays a crucial role here, as the MBZ is topologically equivalent to a torus. One important consequence of this is of course that the integral of Berry curvature over the whole MBZ is quantised in units of the first Chern number $\mathcal{C}_n$. In the above analogy with magnetism, this means that the particle is confined to move on the surface of a torus, while the Chern number counts the number of magnetic monopoles contained inside~\cite{Fang}. 

The above analogy with magnetism is particularly powerful because there are natural limits for our model (\ref{eq:model}) in which the eigenstates of Eq. \ref{eq:mag} and hence of Eq. \ref{eq:dual} are known analytically~\cite{price2014magnetic, ozawa2014momhh, Claassen_prl_2015}. We will focus on the flat-band limit in which the bandwidth is much smaller than the trapping energy; for the energy bands of $\mathcal{H}_0$ with $\alpha = 1/q$, this assumption improves as $\kappa$ decreases or as $q$ increases. In this limit, we can firstly approximate $\Omega_{n}(\mathbf{p}) \approx \Omega_n$, so that the first term is analogous to the kinetic energy of a particle in a uniform magnetic field. Secondly, we can approximate
$E_n(\vt{p}) \approx E_n$ so that the second term of $\widetilde{\mathcal{H}}$ is just a constant energy shift. Hence, the corresponding eigenstates can be understood as toroidal Landau levels in momentum space. We note that the opposite limit, in which the trapping energy is small compared to the bandwidth, also yields very interesting physics including the realisation of a Harper-Hofstatder model in momentum space~\cite{ozawa2014momhh, scaffidi2014exact}. 

As shown in Ref.~\cite{price2014magnetic}, the momentum-space toroidal Landau levels form semi-infinite ladders of states:
\begin{equation}\label{eq:ladders}
  \epsilon_{n,\beta} = E_n  + \left(\beta + \frac{1}{2}\right) \kappa |\Omega_n|  ,
\end{equation}
where we have introduced the Landau level quantum number
$\beta = 0,1,2,\dots$, and where $\kappa |\Omega_n|$ can be recognised as the analogue of the cyclotron frequency
$\omega_c = e |B| /M $. Again, we note that here we have neglected the contribution from the last term in (\ref{eq:dual}), as this will be discussed in the next subsection. As can be shown from the Diophantine equation for the Hall conductivity, for odd values of $q$, the Chern number of all bands except the middle band is $\mathcal{C}_n = -1$~\cite{bernevig2013topological}. Then as the Chern number is related to the uniform Berry curvature as $\mathcal{C}_n = (1/2\pi) \Omega_n A_{\text{MBZ}}$, where $A_{\text{MBZ}} = (2\pi)^2/q$ is the MBZ area, the Berry curvature is given by $|\Omega_n| = \frac{1}{2\pi\alpha}$~\cite{price2014magnetic}.  

While the above spectrum does not directly depend on the toroidal topology of momentum space, the topology does enter into the eigenstate degeneracy, which is equal to $|\mathcal{C}_n|$, as well as into the analytical form of the eigenstates in the MBZ. For example, for the bands with $\mathcal{C}_n = -1$, the eigenstates can be written as~\cite{price2014magnetic}
\begin{eqnarray}\label{eq:chi}
 \chi_\beta (\vt{p}) &=& \mathcal{N}_\beta^{l_{\Omega_n}} \sum_{j = -
 \infty}^{\infty} e^{- i p_y j } e^{ - ( p_x + j  l_{\Omega_n}^2 )
 ^2 / 2 l_{\Omega_n}^2} \nonumber \\ &&
 \times H_\beta ( p_x / l_{\Omega_n} + j 
 l_{\Omega_n})  , \\
 \mathcal{N}_\beta^{l_{\Omega_n}} &=& \left( \frac{\sqrt{2/q}} {2^\beta
\beta! \times 2 \pi l_{\Omega_n}^2} \right)^{1/2} , 
\end{eqnarray}
where $H_\beta$ are the Hermite polynomials and $l_{\Omega_n} = \sqrt{1/|\Omega_n|}$ is the analogue of
the magnetic length. Here we have taken the Berry gauge to have a Landau form $\mathcal{A}_n(\mathbf{p}) = \Omega_n p_x \hat{\vt{p}}_y$ parallel to the $\hat{\vt{p}}_y$ unit vector in  momentum space. We have also assumed that the MBZ is of length $2 \pi$ in one momentum direction, and $2 \pi / q$ in the other direction, corresponding to a magnetic unit cell of $q$ plaquettes containing one flux quantum. While this choice of MBZ is valid in any magnetic gauge of the underlying Harper-Hofstadter model, it is particularly natural when the hopping phases in $\mathcal{H}_0$ are in the Landau magnetic gauge $\phi = (0, 2\pi\alpha m)$, as we shall discuss further below.

\subsection{The Berry connection and the breakdown of approximations}\label{sec:berry-shift}

We now study the effects of the last term in the momentum-space Hamiltonian~(\ref{eq:dual}) which comes from the off-diagonal matrix elements of the Berry connection~\cite{berry}:
\begin{align}
  \delta E_n(\vt{p}) &\equiv \frac{\kappa}{2}\sum_{n^{'}\neq n} \abs{\mathcal{A}_{n,n^{'}}(\vt{p})}^2
  \notag \\
  &=
  \frac{\kappa}{2}\sum_{n^{'}\neq n} \frac{\abs{\bra{u_{n,\vt{p}}}\nabla_{\vt{p}}\mathcal{H}_0(\vt{p})\ket{u_{n^{'},\vt{p}}}}^2}{\left[E_{n^{'}}(\vt{p}) - E_{n}(\vt{p})\right]^2}.
  \label{eq:shift}
\end{align}
This can be recognised as a momentum-space counterpart of the real-space geometrical scalar potential previously studied in atomic systems~\cite{dum:1996, dutta:1999, dalibardrmp2011}. In these systems, the scalar potential arises from real-space Berry connections, which can be created, for example, by using spatially-dependent optical fields to dress the atoms. 
 
In the flat-band limit, we have checked numerically that we can approximate the momentum-space geometrical scalar potential as $\delta E_n(\vt{p}) \approx   \delta E_n$ and so this contributes only a uniform constant energy shift. This term was not considered in our previous works~\cite{price2014magnetic, ozawa2014momhh}, as these works focused on systems such as ultracold atomic gases, where the absolute energy is not easily experimentally observable. As we will discuss in the next section, spectroscopic measurements in photonics are sensitive to the energy of a state, and so such corrections may be extracted experimentally. We note that an analogous effect has also been derived for the effective momentum-space magnetic Hamiltonian for a trapped particle in an ideal flat band~\cite{Claassen_prl_2015}, and predicted for the frequency spectrum associated with excitonic states in transition metal dichalcogenides~\cite{srivastava:2015}.

To see under what conditions the off-diagonal elements of the Berry connection are relevant, we compare the energy $E_{\text{ex}}$ obtained from an exact numerical diagonalization of Eq. \ref{eq:model} with the analytical eigenenergy $  E_{\text{an}}$ predicted by Eq.~\eqref{eq:ladders}. We focus on the lowest ladder of states, associated with band $n=0$ (Eq. \ref{eq:ladders}), at energies which are below the onset of the second ladder around energy $E_1$. This allow us to easily identify which numerical eigenvalue should correspond to which Landau level quantum number~\cite{price2014magnetic}. 

We introduce two dimensionless parameters $\eta_{\text{zpe}}$ and $\eta_{\text{lev}}$ to quantify deviations between the numerics and analytics. The former represents the error in the ``zero-point energy", and is defined as the energy of the lowest numerical state relative to analytical $n=0, \beta=0$ Landau level. The latter is the level spacing error, which we define as the difference between the numerical energy spacing between two neighbouring states and the analytical spacing between states with $\beta$ and $\beta-1$ quantum numbers.  

Considering first the ``zero-point energy" error $\eta_{\text{zpe}}$, the analytical energy of the $n=0, \beta=0$ Landau level is given from Eq. \ref{eq:ladders} by:
\begin{equation}
  E_{\text{an}} = \left<E_0(\vt{p})\right>_{\vt{p}} + \frac{1}{2}\frac{\kappa}{2\pi\alpha} ,
\end{equation}
where we have used that $|\Omega_n| = \frac{1}{2\pi\alpha}$ and where we calculate the uniform energy shift $E_0=\left<E_0(\vt{p})\right>_{\vt{p}}$ as the average band-energy over the MBZ. This definition generalises our flat-band approximation to account for the non-zero bandwidth of the lowest band. 

\begin{figure}[tb]\centering
  \includegraphics[width=0.9\linewidth]{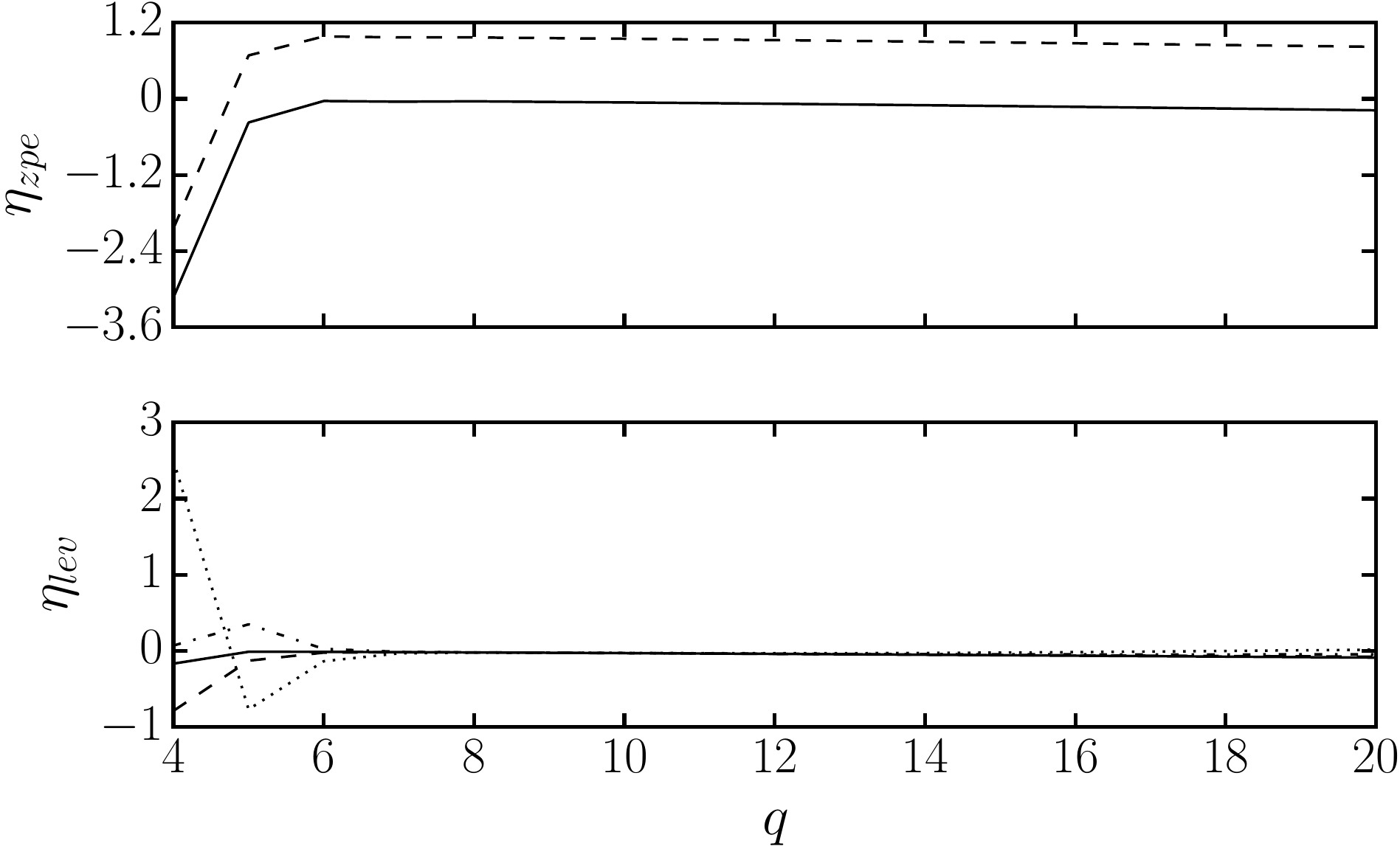} % anc/scripts/nonabelian_fig.jl
  \caption{\emph{Top panel}: ``Zero-point energy" error, with (solid curve,
    $\eta_{\text{zpe}}^{\text{nab}}$) and without (dashed line,
    $\eta_{\text{zpe}}$) the shift from the off-diagonal matrix elements of the Berry connection. Including just the
    first term in the sum Eq.~\eqref{eq:shift} gives an identical
    curve to the one of $\eta_{\text{zpe}}^{\text{nab}}$. The chosen
    trap strength is $\kappa = 0.02 J$.  \emph{Bottom panel}: Level-spacing
    error, for the same trap strength, considering $\beta = 0,1$
    (solid curve), $\beta = 1,2$ (dashed curve), $\beta = 2,3$
    (dashed-dotted curve) and $\beta = 3,4$ (dotted curve).}
  \label{fig:zpe}
\end{figure}

We then define the dimensionless parameter $\eta_{\text{zpe}}$ as 
\begin{equation}
\eta_{\text{zpe}} = \frac{4\pi\alpha}{\kappa} (E_{\text{ex}} -E_{\text{an}}). 
\end{equation}
This dimensionless error is plotted with a dashed line as a function of $q$ for $\kappa=0.02 J$ in the top panel of Fig.~\ref{fig:zpe}. At small $q$, there is a large bandgap between the lowest two Harper-Hofstadter energy bands, but the lowest band also has a large bandwidth. In this regime, the single-band approximation is reasonable, while the flat-band approximation breaks down leading to large errors. This limit requires a different analytical approach as  previously presented in Ref.~\onlinecite{ozawa2014momhh}.
To account for the error at large $q$, we include the shift from the off-diagonal matrix elements of the Berry connection (Eq. \ref{eq:shift}). We calculate this shift numerically from the eigenstates of the Harper-Hofstadter model, and we incorporate it into a second parameter 
\begin{equation}
\eta_{\text{zpe}}^{\text{nab}} = \frac{4\pi\alpha}{\kappa}(E_{\text{ex}} - E_{\text{an}} - \left<\delta  E_0(\vt{p})\right>_{\vt{p}}).
\end{equation}
This is plotted as a solid line in the top panel of Fig.~\ref{fig:zpe}. As can be seen, this shift dramatically reduces the error in the zero point energy at large $q$. We also calculate this shift considering only the effects of band mixing with the second lowest band $n=1$; this is indistinguishable on this scale from the full shift. This can be understood from the dependence on the bandgaps in Eq.~\ref{eq:shift}, which shows that the contributions of high-energy bands are suppressed. As discussed further in Section \ref{sec:experiment}, it would be possible experimentally to extract the energy of the lowest state; this could constitute the first direct measurement of the effects of the off-diagonal matrix elements of the Berry connection in a photonics system.

We turn now to the level-spacing error $\eta_{\text{lev}}$. This can be expressed as 
\begin{equation}
\eta_{\text{lev}} = \frac{2\pi \alpha}{\kappa} [E_{\text{ex}}(\beta)
- E_{\text{ex}}(\beta - 1)] -1, 
\end{equation}
where we have used that the analytical level spacing from Eq. \ref{eq:ladders} is simply $\kappa / 2\pi \alpha$. 
We plot the level-spacing error in the bottom panel of Fig.~\ref{fig:zpe} for $\beta = 1, 2, 3$ and 4. As can be seen here, there is a large variation in the errors at small $q$ due to the large bandwidth~\cite{ozawa2014momhh}. On the other hand, we see that $\eta_{\text{lev}} \ll 1$, for $q \gtrsim 6$, where the flat-band approximation improves. In this regime, the level-spacing error is much smaller than the zero-point error. This is because when the shift from the off-diagonal matrix elements of the Berry connection (Eq. \ref{eq:shift}) is approximately uniform over the MBZ at large $q$, it just acts as a uniform energy shift on all the states in a  ladder with band index $n$. Consequently, this shift drops out of the level spacing error between states, leaving only higher-order band-mixing terms. From perturbation theory, it is expected that mixing with other bands leads to a negative energy shift on states in the lowest band, and indeed this can be seen in both $\eta_{\text{zpe}}^{\text{nab}}$ and $\eta_{\text{lev}}$ in the small negative errors found at large $q$. 

\subsection{Driving and dissipation}\label{sec:driven-dissipation}

We now include in our model the driving and dissipation that are an integral part of the proposed photonics experiment. We assume there are uniform and local losses
characterized by a loss rate $\gamma$, and that the pump is
monochromatic, with frequency $\omega_0$ and a spatial profile
$f_{m,n}$. Following the treatment of Ref.~\onlinecite{ozawa2014qhe}, we replace the bosonic creation and annihilation operators with their expectation
values, as can be justified for a noninteracting system. The steady state evolution of the photon-field amplitude in a cavity then follows that of
the pump as $a_{m,n}(t) = a_{m,n} e^{-i \omega_0 t}$. Combining Hamiltonian evolution with pumping and losses, one arrives at a set of linear coupled equations that can be solved numerically for the steady-state\cite{cohen1992atom}:
\begin{eqnarray}\label{eq:linear_problem}
f_{m,n} =J&&\left[e^{-i\phi_{m,n}^x}a_{m+1,n}+e^{i\phi_{m-1,n}^x}a_{m-1,n} \right. \nonumber\\ &&\left. +e^{-i\phi_{m,n}^y}a_{m,n+1}+e^{i\phi_{m,n-1}^y}a_{m,n-1}\right] \nonumber\\
&&{\resizebox{.8\hsize}{!}{${+\left[\omega_{0}+i\gamma-\frac{1}{2}\kappa 
\left((m-m_0)^{2}+(n-n_0)^{2}\right)\right]a_{m,n}}$}} \qquad
\end{eqnarray}
where we have reintroduced the position of the harmonic trap centre $(m_0, n_0)$, although unless otherwise specified we set $(m_0, n_0)=(0,0)$ in our simulations.

The expectation values $\abs{a_{m,n}}^2$ correspond to the number of photons
at site $(m,n)$, whereas the intensity spectrum is given by their
total sum $\sum_{m,n} |a_{m,n}|^2$ as a function of pump frequency
$\omega_0$. These observables can be directly related to the eigenstates of the Hamiltonian in Eq. \ref{eq:model}. Firstly, the different eigenmodes of a driven-dissipative system will appear as peaks in the transmission and/or absorption spectra under a coherent pump~\cite{carusotto2013fluids}. The resonance peaks will be broadened by the decay rate $\gamma$, while the area of the peaks will depend on the overlap between the spatial amplitude
profile of the pump and the underlying eigenstate of $\mathcal{H}$ at that energy. 

Secondly, when the pump frequency is set on resonance with a given mode, the intensity profiles in both real- and momentum-space reproduce the wave function of that mode~\cite{carusotto2013fluids}. This corresponds respectively to measuring the near-field and far-field spatial emission of photons from the cavity array. We note that the far-field emission is simply the Fourier-transform of the real-space wave function and so will be a function of crystal momentum defined in the full BZ. To reach the MBZ, a  further processing step is required; for example, if the Harper-Hofstadter Hamiltonian is in the Landau gauge and if we choose a magnetic unit cell of $q$ plaquettes along $\hat{x}$, the appropriate transformation takes a particularly simple form~\cite{price2014magnetic}:
\begin{equation} \label{eq:trans}
  \sum_n \abs{\psi_n(\vt{p}_\mathrm{MBZ})}^2 = \sum_j \abs{\psi(\vt{p}_\mathrm{BZ} = \vt{p}_\mathrm{MBZ}- j\vt{G})}^2 ,
\end{equation}
where $\psi_n(\vt{p}_\mathrm{MBZ})$ is the wave function coefficient in the MBZ, while $\psi(\vt{p}_\mathrm{BZ})$ is that in the original BZ. In this expression, $j$ is an integer, while $\vt{G} = (2\pi/q) \hat{\vt{p}}_x $ is the magnetic reciprocal lattice vector, where the factor of $q$ is due to the enlarged magnetic unit cell. We note that for other magnetic gauges or for other choices of the magnetic unit cell, this transformation will in general be more complicated. In this sense, we call this choice of magnetic unit cell, a ``natural" choice when the Harper-Hofstadter Hamiltonian is in the Landau gauge.
In the rest of the article, we denote the momentum in the original BZ as $\mathbf{p}$, and that in the MBZ as $\mathbf{p}^0$.

\section{Results and discussion}
\label{sec:results}

\subsection{Pumping \& gauge-dependent effects}
\label{sec:selection}

\begin{figure}[tb]\centering
  \includegraphics[width=\linewidth]{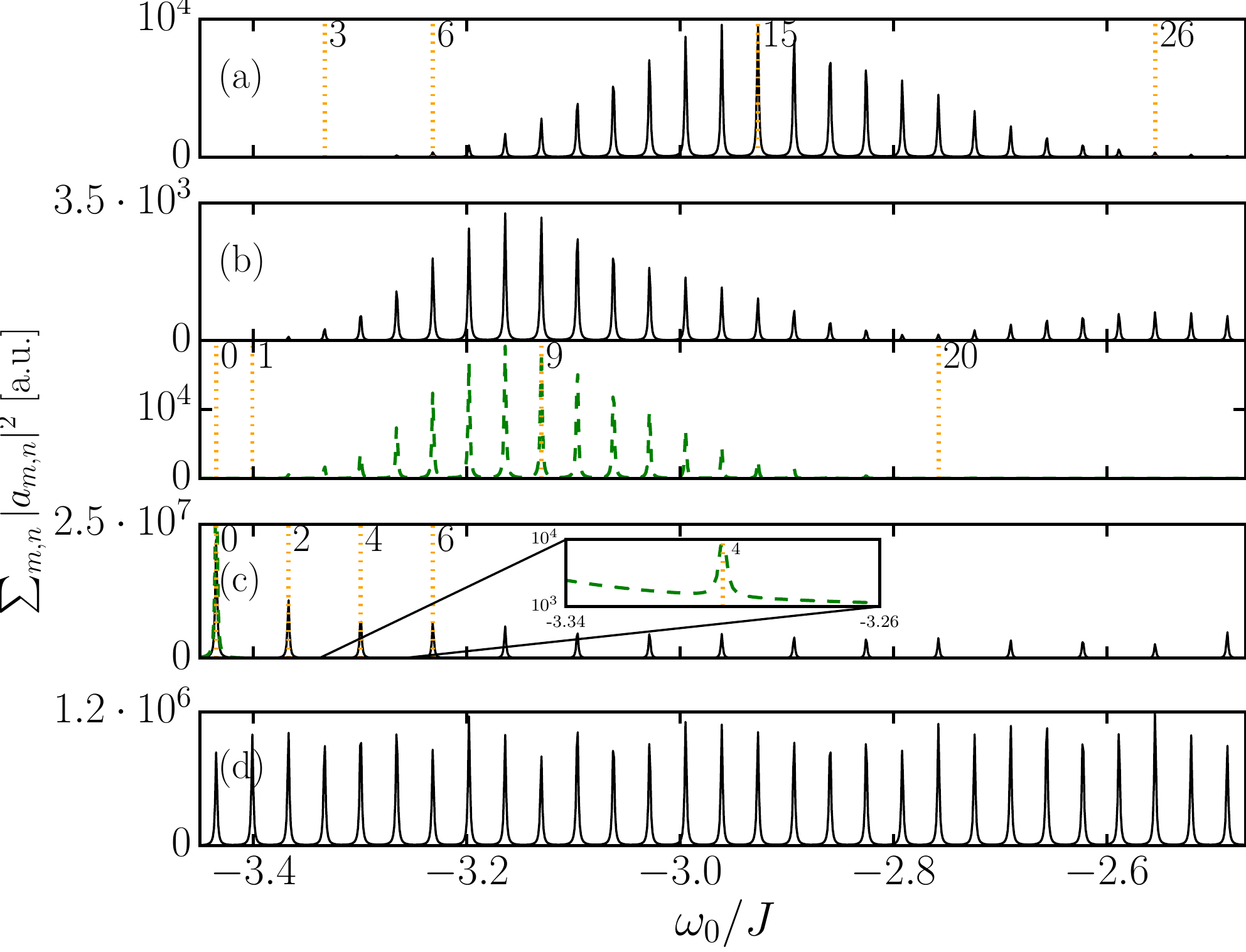} % anc/scripts/selection_fig.jl
  \caption{(Color online) Intensity spectra for different pumping conditions: (a)
    pumping the single site (5,5), (b) pumping with a Gaussian
    profile centered at site (5,5) with width $\sigma=1$, (c) homogeneous pumping across all lattice sites and (d) pumping with a random
    phase across all lattice sites. These results were obtained by numerical solving Eq.~\eqref{eq:linear_problem} for the steady-state in a lattice of
$N \times N = 45 \times 45$ sites, with $\kappa = 0.02 J$,
$\gamma = 0.001 J$ and $\alpha = 1/11$.   
    Black (solid) curves correspond to using the Landau gauge while
    green (dashed) ones to the symmetric gauge. The dotted vertical
    lines (with labels indicating the value of $\beta$) mark the
    states which were selected for later analysis. The spectra in
    panels (a) and (d) are identical for both gauges.}
  \label{fig:pumping_schemes}
\end{figure}

As introduced above, spectroscopic measurements can be used in a driven-dissipative photonics experiment to study the trapped Harper-Hofstadter model and hence toroidal Landau levels in momentum space. In this section, we focus on the effects of the pumping, exploring how different pumping schemes excite the eigenstates with different weights. We find that such spectroscopic measurements are sensitive to the underlying synthetic magnetic gauge chosen in a given implementation of the Harper-Hofstadter Hamiltonian. 

To best illustrate these gauge-dependent effects, we present the results of numerically solving
Eq.~\eqref{eq:linear_problem} for the steady-state in a large lattice of
$N \times N = 45 \times 45$ sites, with $\kappa = 0.02 J$,
$\gamma = 0.001 J$ and $\alpha = 1/11$.  These parameters are chosen to highlight the key features of different pumping schemes; we will present numerical results for a more realistic experimental system in Section \ref{sec:experiment}.  The numerical code was written
in \textsc{Julia}~\cite{bezanson2014julia} and is available in the Supplemental Material~\footnote{See Supplemental Material.}. 

The intensity spectrum of the steady-state as a function of pump
frequency is shown in Fig.~\ref{fig:pumping_schemes}, where we compare results for both the Landau and symmetric gauge for four pumping
schemes, discussed in turn below. For simplicity we limit ourselves to pump frequencies located between the two lowest-lying Harper-Hofstadter bands of the untrapped system. This allows us to focus only on states within the first ladder of the trapped system (Eq.~\eqref{eq:ladders} with $n = 0$). At higher energies, the clear identification of states is more difficult as more than one ladder of toroidal Landau levels can overlap, as shown, for example, in Section \ref{sec:experiment}. 

{\em{Single-site pumping--}}The first and simplest case that we consider is that of pumping a single site $f_{m,n} = \delta_{m,m_0} \delta_{n,n_0}$ at an off-center lattice site. These results are shown in Fig.~\ref{fig:pumping_schemes} panel (a), where the uniform energy spacing of the toroidal Landau levels can be clearly observed. For this pumping scheme, we find no significant differences between the spectra for the Landau or symmetric gauge. This is to be expected as changing the gauge is equivalent
to changing the relative phase between different sites, but as we are only pumping one site, this phase difference is unimportant.

\begin{figure}[tb]
  \centering
  \includegraphics[width=0.9\linewidth]{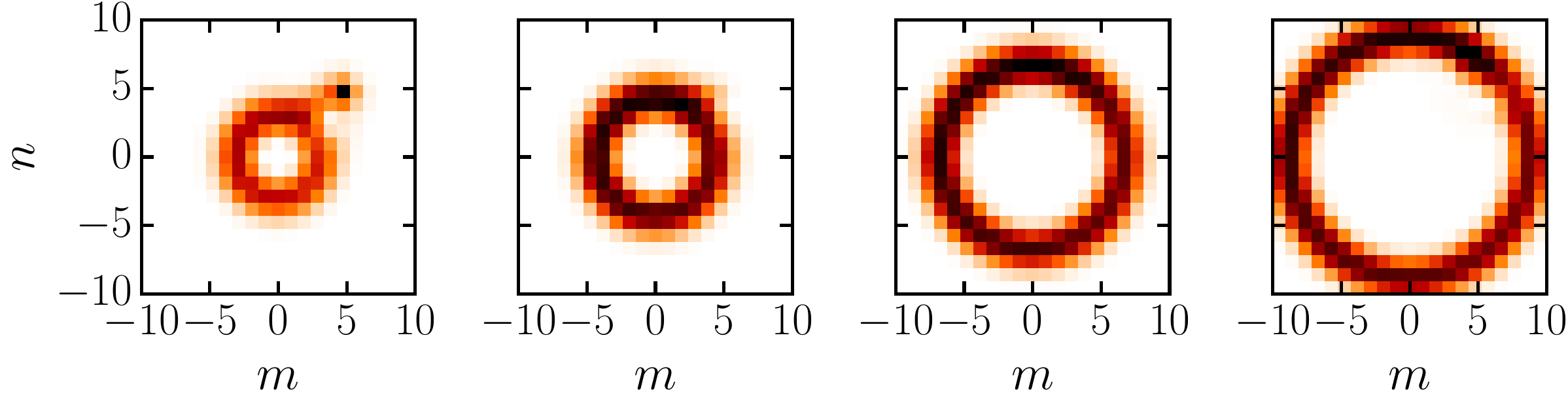} % anc/scripts/selection_fig.jl
  \caption{(Color online) Real space reconstruction of the states $\beta=3$, 6, 15
    and 26 of Fig.~\ref{fig:pumping_schemes}(a).}
  \label{fig:delta_real_sp}
\end{figure}

 Instead, for both magnetic gauges, we see that the peak height is very small for low energy states, rising to a maximum as energy increases, before decreasing once more. This behaviour can be understand by considering the form of the real space wavefunctions of $\mathcal{H}$. 
In real space, the eigenstates are rings of finite width which increase in radius as the energy increases (as can be seen in Fig.~\ref{fig:delta_real_sp}). Analytically, we can predict how the ring radius scales with energy by remembering that the term
$(\beta + 1/2) \kappa |\Omega_n|$ in Eq. \ref{eq:ladders} is the momentum-space kinetic energy
$\frac{\kappa}{2}r^2$ where $r = i\nabla_{\mathbf{p}} + \mathcal{A}_{0, 0}(\mathbf{p})$ is the physical position operator in the lowest band. From this, we deduce that $r^2 \approx \frac{1}{\pi} q \beta$, as can be confirmed numerically. Therefore, if one pumps an off-center site, there will only be a
limited range of rings that will have radii that will overlap with
the pump spot and so be excited. Here, we have set the pump spot to be at position $(5,5)$, and from the above scaling, the toroidal Landau level that best overlaps with this pump will have a quantum number $\beta \approx 14$, which is in good agreement with the numerical results shown in
Fig.~\ref{fig:pumping_schemes} (a). 

{\em{Gaussian pumping--}} We now consider a Gaussian pump as the next logical step up in complexity from a single-site pump. This has the form $f_{m,n} = \exp- \frac{1}{2\sigma^2} \left[(m-m')^2 + (n-n')^2
\right]$, and we choose $\sigma =1$ and for the pump centre to be at $(m',n') = (5,5)$, as for single-site pumping. The results are shown in Fig.~\ref{fig:pumping_schemes} (b). The main effect is, as expected, that more states
become visible in both the low- (smaller $\beta$) and high-energy (larger $\beta$) sections of the spectrum. This is because the pump has a greater spatial width and so overlaps with a larger range of real-space eigenstates. 

\begin{figure}[tb]\centering
  \includegraphics[width=0.9\linewidth]{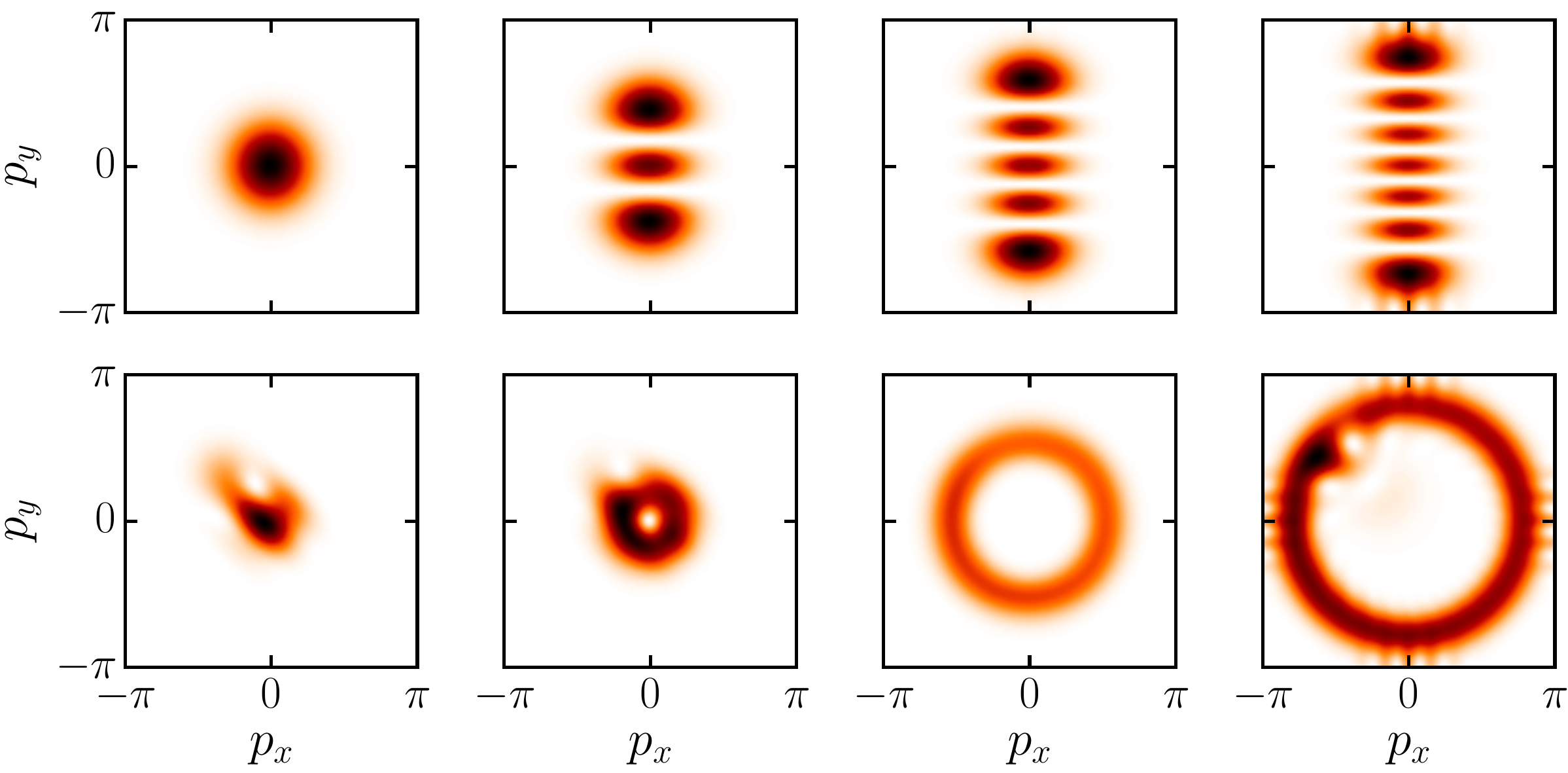} % anc/scripts/selection_fig.jl
  \caption{(Color online) Momentum space reconstruction of the
    eigenstates. \emph{Top row}: states corresponding to $\beta=0$, 2,
    4 and 6 in Fig.~\ref{fig:pumping_schemes}(c), using the
    Landau gauge.  \emph{Bottom row}: states corresponding to
    $\beta=0$, 1, 9 and 20 in Fig.~\ref{fig:pumping_schemes}(b), using the symmetric gauge.}
  \label{fig:hom_mom_sp}
\end{figure}

However, we can also see that the intensity spectrum now depends on the underlying magnetic gauge, as the phase of the eigenstates is important. In particular, more high-energy peaks can be seen for the Landau gauge than for the symmetric gauge. This can be most easily understood by noting that in momentum space, the symmetric-gauge states also have a ring-like structure (see bottom panel of Fig.~\ref{fig:hom_mom_sp}), where the ring radius increases with $\beta$. To see this, we note that, in the symmetric gauge, the real-space wavefunctions have a phase which
winds around the ring as $e^{i\beta \phi}$ where $\phi$ is the polar angle around the ring. This phase-winding sets the radius of the rings in momentum space as $p^2 \approx \pi \frac{\beta}{q}$; a scaling that can be confirmed numerically and seen in Fig.~\ref{fig:hom_mom_sp}, bottom panel. (The white spot close to the edges of the rings in these figures is due to destructive interference with the pump.)  As the Fourier transform of the Gaussian pump is again a Gaussian, it follows that only a limited range of low-energy symmetric-gauge momentum-space states will have a good overlap with the pump. The high energy portion of the spectrum is therefore washed out compared to its Landau gauge
counterpart, where states have higher amplitude close to the centre of the BZ and so better overlap with the pump. 

\begin{figure}[tb]
  \centering
  \includegraphics[width=0.9\linewidth]{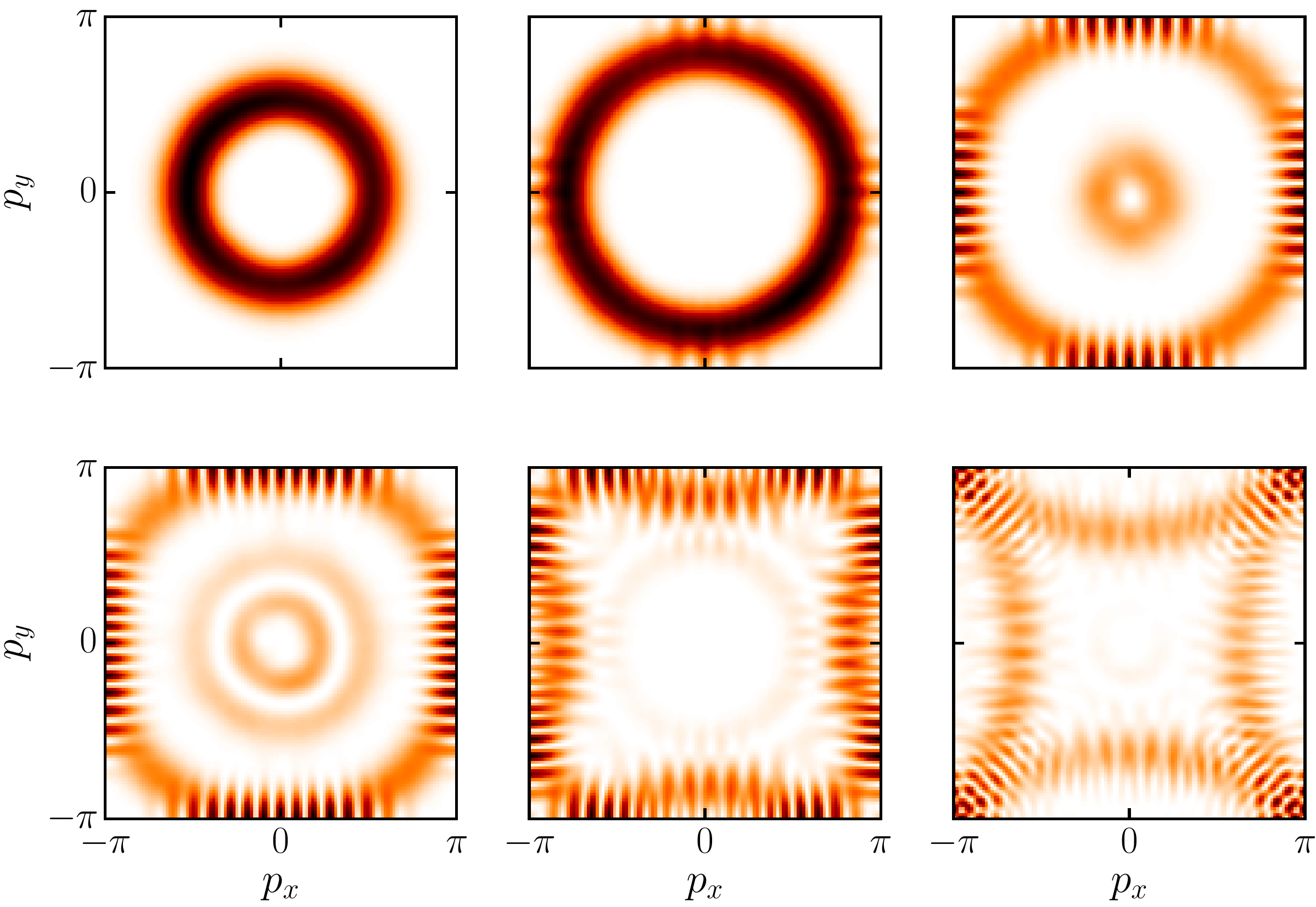} % anc/scripts/torus_edge_fig.jl
  \caption{(Color online) Momentum space reconstruction of the eigenstates in the
    full BZ, using the symmetric gauge and homogeneous pumping with a random on-site
    phase. \emph{Top row}: states corresponding to
    $\beta = 9, 20, 30$.  \emph{Bottom row}: states corresponding to
    $\beta = 38, 59, 99$. Parameters are the same as in Fig.~\ref{fig:pumping_schemes}.}
  \label{fig:torus_edge}
\end{figure}

Before continuing, we also note that for sufficiently large values of $\beta$ the symmetric-gauge rings in momentum space will increase to the point where they touch the BZ boundaries. When this occurs, self-interference patterns appear in the wave function as shown for example in Fig.~\ref{fig:torus_edge}. The extra ring-like structures appearing for $\beta \geq 30$ in Fig.~\ref{fig:torus_edge} are due to the close proximity of states pertaining to other ladders with
$n >0$. Note that in order to excite such high-energy states, we have used a pump with a homogeneous amplitude and a random onsite phase, as will be presented as the fourth pumping case below. 

{\em{Homogeneous pumping with uniform phase--}} If we now take the limit of a very wide Gaussian, we reach a homogeneous pump profile extended over all lattice sites. The results for this pumping scheme are shown in panel (c) of
Fig.~\ref{fig:pumping_schemes}. Now $f_{m,n} = f$, and we see that the intensity spectrum is strongly magnetic-gauge dependent. In the Landau gauge, firstly, there are visible peaks for only
half of the states. This can be understood by noting that a homogeneous pump in real space is a $\delta$ function in momentum space centered in the
middle of the BZ. If we consider the Landau-gauge eigenstates in the full BZ, as shown
in the top row of Fig.~\ref{fig:hom_mom_sp}, we see that the states with an even
value of $\beta$ have an even number of nodes, with a lobe at the BZ
center. Conversely, the states with odd values of $\beta$ have an odd number of nodes, including one at the BZ center. (These feature can be related back to the properties of the Hermite polynomials in the analytical eigenstates in the MBZ (Eq.~\eqref{eq:chi}).) Consequently, only states with even values of $\beta$ have a good overlap with the pump, and the intensity spectrum contains half the expected peaks, now separated by twice the toroidal Landau level energy spacing.  

In the symmetric gauge, secondly, we find only one out of every four
states for homogeneous pumping, as can be seen in the inset of Fig.~\ref{fig:pumping_schemes}
(c). This is due to the fact that, on a square lattice, the angular
momentum is conserved modulo 4, respecting the 4-fold rotational
symmetry. The peak intensity gets smaller for larger $\beta$ because
of the diminishing overlap of the localized central pump with the
increasing momentum-space ring discussed above.

{\em{Homogeneous pumping with a random on-site phase --}} As the fourth scheme, we consider a pump with a uniform amplitude over the lattice but a random site-dependent phase $\phi_{m,n}$:
$f_{m,n}=fe^{i\phi_{m,n}}$. The phases are chosen from a random
uniform distribution, and have values in the interval $[0,2\pi)$. The
bottom panel of Fig.~\ref{fig:pumping_schemes} was obtained by
averaging over 100 distinct realizations of these random phases. This
results in a relatively even intensity distribution, for both gauges, where we can associate a peak to each toroidal Landau level in this energy window. While such a pumping scheme would therefore be the best way to excite all the eigenstates and to fully probe the momentum-space physics, we note that this would also be difficult to achieve in an experiment.

\begin{figure}[htb]
  \centering
  \includegraphics[width=\linewidth]{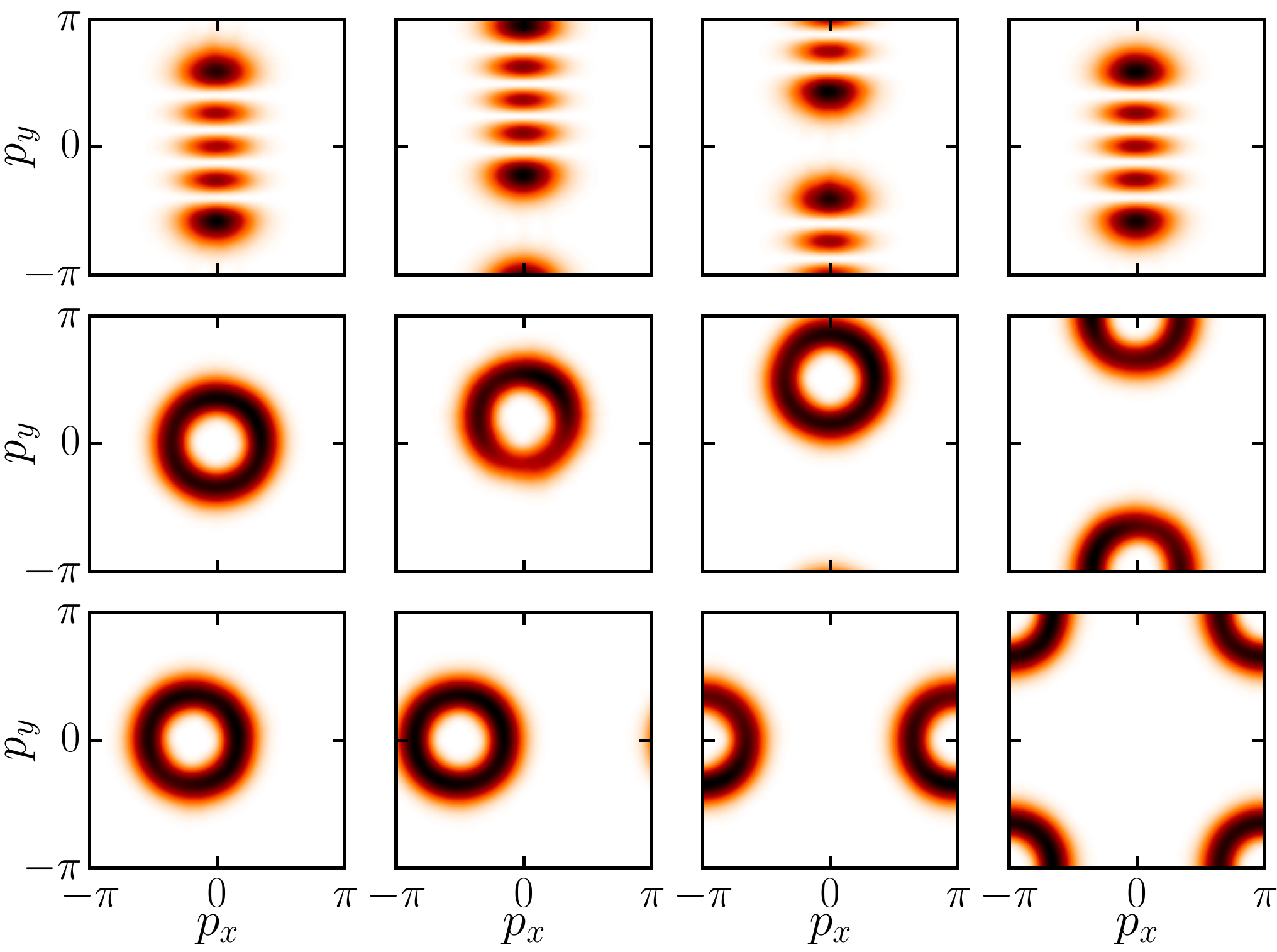} % anc/scripts/torus_edge_fig.jl
  \caption{(Color online) Momentum space reconstruction of the eigenstates in the
    full BZ, using the Landau (top row) and symmetric gauge (middle
    and bottom row) and a spatially homogeneous pump with
    a random on-site phase. We have considered the state $\beta = 4$ for
    different trap positions $(m_0, n_0)$. For the top and middle
    rows, we have (from left to right): (0,0) (trap in the center),
    (2,0), (5.5,0) and (11,0), whereas for the bottom row we chose the
    positions (0,2), (0,5.5), (0,11) and (11,11). Parameters are the same as in Fig.~\ref{fig:pumping_schemes}.}
  \label{fig:moving_trap}
\end{figure}

Before continuing, we give a final example of an interesting gauge-dependent effect that could be studied experimentally in this system. Unlike the physics discussed above, this is not directly related to the pumping but instead to the behaviour of the wave function under a change in the centre of the harmonic trap $(m_0, n_0)$. As derived in Ref. ~\onlinecite{ozawa2014momhh}, moving the
harmonic trap in space changes the boundary conditions on the wave function in the MBZ. We note that although this derivation was made explicitly for the magnetic Landau gauge in the MBZ, numerically we observe here that this physics is also observed in the full BZ in both gauges. 
As shown in Fig.~\ref{fig:moving_trap}, a shift in the harmonic trap centre in one direction shifts the observed momentum-space pattern in the perpendicular direction. This behaviour can be understood as a realisation of Laughlin's {\em Gedankenexperiment} for the quantum Hall effect but now in momentum space~\cite{ozawa2014momhh}. As we observe, the momentum-space wave function returns to itself after the harmonic trap has been moved $q$ lattice sites for the magnetic Landau gauge but $2q$ lattice sites for the magnetic symmetric gauge, reflecting the underlying translational symmetry of $\mathcal{H}_0$ in the two different gauges.

\subsection{Results for realistic experimental parameters}
\label{sec:experiment}

We now present numerical results for system parameters within current experimental reach, to demonstrate that the essential characteristics of toroidal Landau levels could be probed experimentally for the first time in photonics. We choose a small
lattice of only $11 \times 11$ sites, with losses of $\gamma = 0.05 J$; this loss rate is in the same range as those present in the experiment of
Ref.~\onlinecite{hafezi2013imaging}. Such a large loss rate broadens the
peaks in the intensity spectrum, making closely-spaced eigenenergies harder to resolve. From
Eq.~\eqref{eq:ladders}, we see that the level spacing is given by
$\frac{\kappa}{2\pi\alpha}$, and so we can increase the energy spacing by applying a stronger harmonic potential, chosen here as $\kappa = 0.2 J$. Increasing the strength of the harmonic trap improves our flat-band approximation, but weakens the single-band approximation. To compensate for this, we consider a larger value of $\alpha = 1/7$, for which the larger band-gap $(E_1 - E_0)$ reduces band-mixing effects. 

\begin{figure}[tb]
  \centering
  \includegraphics[width=\linewidth]{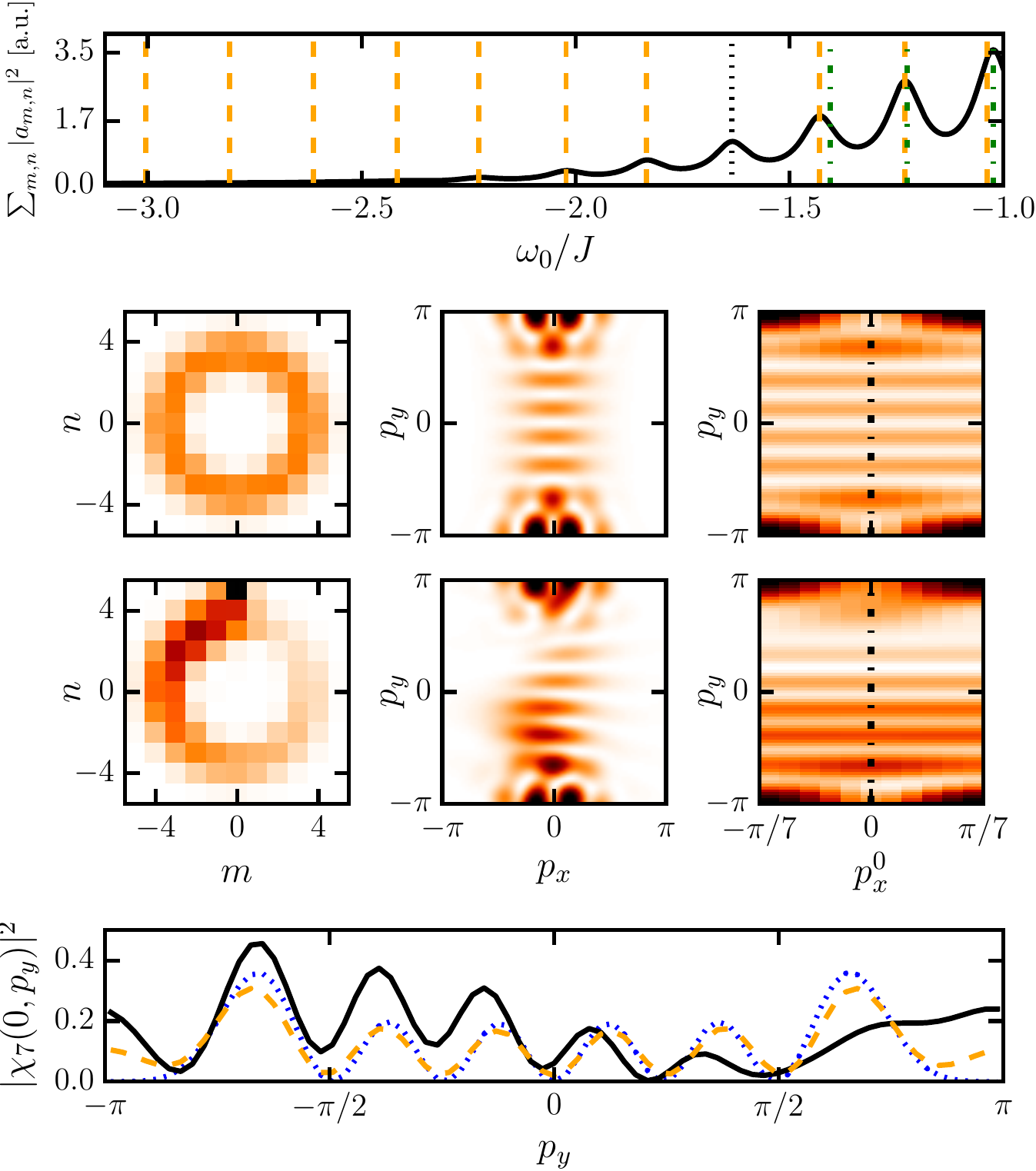} % anc/scripts/experimental_fig.jl
  \caption{(Color online) \emph{Top row}: Intensity spectrum for a small lattice of $11 \times 11$ sites, with
    $\gamma = 0.05 J$, $\kappa = 0.2 J$ and $\alpha=1/7$. The orange vertical (dashed) lines
    show the first ladder of eigenstates of $\mathcal{H}$ from Eq.~\eqref{eq:model}. The second ladder of states is indicated with green vertical dash-dotted lines.
  \emph{Second row}: Profile of the $\beta=7$ state in real space
    (left), momentum space (center) and the population over bands in the MBZ (right) for the conservative system with no pumping nor decay.
    \emph{Third row}: Reconstruction of the $\beta=7$ mode wavefunction in real space (left), momentum space (center) and the population over bands in the MBZ (right) obtained in the driven-dissipative system by setting the pump frequency at $\omega_0 = -1.63 J$ on resonance with the desired mode (black dotted line in the top panel).
The $\delta$-like pump at (0,5) is visible as a dark square in the left panel.
    \emph{Bottom row}: Slice along the $p_x^0 = 0$ line in the MBZ
    (solid black line) compared to the analytical prediction of
    Eq.~\eqref{eq:chi}, $|\chi_7(0,p_y)|^2$ (blue dotted line) and to the population over bands for the nondissipative system (dashed orange line).}
  \label{fig:exp}
\end{figure}

As in the experiment of Ref.~\onlinecite{hafezi2013imaging}, we work in the Landau gauge for the Harper-Hofstadter Hamiltonian, with hopping phases given by $\phi = (0, 2\pi\alpha m)$. In order to model the experimental pumping scheme where light was injected into a single resonator at the edge of the system via an external integrated waveguide~\cite{hafezi2013imaging}, we consider a localized pump on a single site situated on the upper border of the system at $(m_0,n_0)= (0,5)$. 
The corresponding intensity spectrum is shown in the 1st row of Fig.~\ref{fig:exp}. Apart from the expected
broadening due to larger losses, the peaks observed correspond well to the expected eigenenergies. 
As discussed above, single-site pumping limits the number of visible peaks, as the heights of the peaks at low-energies are suppressed due to the poor overlap of the real-space eigenstate with the pump position. However, as this pumping scheme is closest to that used in experiments, we emphasise that even in this case, enough peaks can be observed to extract quantitative measurements of the toroidal Landau level spacing. We also note that here for frequencies larger
than $-1.5 J$, we also start to see states from the second ladder
$\epsilon_{1,\beta}$ (see Eq.~\eqref{eq:ladders}), which are depicted as green
vertical dash-dotted lines. Their proximity to the first ladder states
means they cannot be easily resolved as separate peaks in the dissipative
spectrum.

Setting the pump frequency at the energy indicated by the black dotted
line, we plot the numerical near- and far-field emission in the left and center panels of the 3rd row of Fig.~\ref{fig:exp}. This corresponds to 
the wave function in real space and in the full
BZ respectively. By applying the transformation in Eq. \ref{eq:trans}, we can also map the wave function in the full BZ to the population over bands in the MBZ, as shown in the right panel of the 3rd row of Fig.~\ref{fig:exp}. For comparison, we plot these
quantities for the corresponding numerical eigenstate of $\mathcal{H}$ (Eq.~\eqref{eq:model})
in the second row of Fig.~\ref{fig:exp}, for which there is no pumping and dissipation. 

We find very good qualitative agreement between the numerical results in the MBZ and the analytical toroidal Landau level (\ref{eq:chi}) with $\beta=7$ as expected. We can make a quantitative comparison with this analytical eigenstate by taking
a slice along the dash-dotted vertical lines ($p_x^0 = 0$) in the
right column of rows 2 and 3; these cuts are shown in the bottom panel of Fig.~\ref{fig:exp} along with a dotted blue curve indicating the analytical eigenstate. As can be seen, there is excellent agreement between the numerics without driving and dissipation and the analytical result. We have checked that reducing $\kappa$ makes this fit even better, pointing towards band-mixing effects. Introducing pumping and dissipation distorts the eigenstate, but many characteristic features are still clearly observable. 

It is particularly interesting to note that in the driven-dissipative steady-state in real space, shown in the left panel of the 3rd row of Fig.~\ref{fig:exp}, the photon distribution breaks the rotational symmetry of the ring eigenstate. While this can be physically understood 
as a
decaying cyclotron orbit with an inverse lifetime set by $\gamma$, in terms of eigenmodes the exponential decay (and more generally the breaking of the rotational symmetry) results from the interference of several modes which overlap in frequency due to the relatively large value of $\gamma$.
In the same way that real-space Landau
levels give rise to real-space cyclotron orbits under the effect of the magnetic field, the observation of momentum-space
Landau levels can provide clear evidence of a cyclotron orbit in momentum space under the effect of the Berry curvature, whose effect is indeed that of a momentum-space magnetic field.

Finally, we briefly summarize how one can practically measure the
contribution $\delta E_0$ from the off-diagonal matrix elements of the Berry connection
 (see Eq.~\eqref{eq:shift}) from the intensity spectrum. Starting from an experimental
spectrum, one first needs to select a particular peak and determine
its $\beta$ label by comparing the MBZ reconstruction with the
analytical result. The distance between two neighbouring peaks gives
the level spacing $\kappa \abs{\Omega_0}$. Finally, to separate the shift $\delta E_0$ from the Harper-Hofstadter ground state
energy $E_0$ in Eq.~\eqref{eq:ladders}, one can make use of the fact
that the former depends on the trap strength $\kappa$, while the
latter does not. Preparing two otherwise identical samples with different trap
strengths and subtracting the ground state energy will then allow for a
direct measurement of the contribution from the off-diagonal matrix elements of the Berry connection. 

\section{Conclusion}
\label{sec:conclusion}

In conclusion, we have shown that the observation of toroidal Landau levels in momentum
space is within experimental reach for state-of-the-art driven-dissipative photonic
systems. Our proposal combines the recent realisation of the Harper-Hofstatder model in an array of silicon-based coupled ring resonators in
Ref.~\onlinecite{hafezi2013imaging}, with a harmonic potential, which could be introduced through a spatial modulation of the resonator size. We have presented numerical results to show that even for very small lattices, in the presence of driving and strong dissipation, key characteristics of the toroidal Landau levels can still be extracted. This would be a first direct investigation of analogue magnetic eigenstates in momentum space. 

We have also emphasised that the proposed photonics experiment would be able to highlight a momentum-space analog of the cyclotron motion as well as to measure the energy shift due to the off-diagonal matrix elements of the Berry connection, which, as these are inter-band geometrical properties, are hard to access by other means. We have also discussed how the spectroscopic measurements presented here are sensitive to the specific synthetic magnetic gauge implemented in an experiment. 

Finally, an interesting outlook would be to include the effect of photon-photon interactions in the model, as the degenerate ground states predicted in~\cite{ozawa2014momhh} for a weakly-interacting trapped Harper-Hofstadter model may lead to interesting nonlinear dynamical features. In the longer run, when the synthetic gauge field is combined with strong interactions, one can hope to observe the hallmarks of fractional quantum Hall physics~\cite{umucalilar2012fractional,hafezi2013non}.

\acknowledgments

We are grateful to Ajit Srivastava, Ata\c{c} Imamo\v{g}lu and Germain Rousseaux for stimulating discussions. A.C.B. acknowledges financial support from the ESF through the POLATOM grant 4914. H.M.P., T.O. and I.C. were funded by ERC through the QGBE grant, by the EU-FET Proactive grant AQuS (Project No. 640800), and by Provincia Autonoma di Trento, partially through the project ``On silicon chip quantum optics for quantum computing and secure communications - SiQuro". H.M.P was also supported by the EC through the H2020 Marie Sklodowska-Curie Action, Individual Fellowship Grant No: 656093 ``SynOptic".

\end{document}